\documentclass[namedreferences]{solarphysics}

\usepackage[hyperref,optionalrh]{spr-sola-addons}
\usepackage{graphicx}        
\usepackage{color}           
\usepackage{breakurl}        

\chardef\us=`\_

\begin{document}

\begin{article}
\begin{opening}

\title{A synchronized two-dimensional $\alpha-\Omega$ model of the solar dynamo }


\author[addressref={aff1,aff2}]{\inits{M.}\fnm{M.}~\lnm{Klevs}}\sep
\author[addressref={aff1},corref,email={F.Stefani@hzdr.de}]{\inits{F.}\fnm{F.}~\lnm{Stefani}}\sep
\author[addressref={aff3}]{\inits{L.}\fnm{L.}~\lnm{Jouve}}\sep
\address[id=aff1]{Helmholtz-Zentrum Dresden -- Rossendorf, Bautzner Landstr. 400,
D-01328 Dresden, Germany}
\address[id=aff2]{University of Latvia, Institute for Numerical Modelling, 3 Jelgavas street, 
Riga, LV-1004, Latvia}
\address[id=aff3]{Univ. Toulouse, IRAP, CNRS, UMR 5277,CNES,UPS, F-31400 Toulouse, France}

\runningauthor{M. Klevs {\it et al.}}
\runningtitle{A synchronized two-dimensional $\alpha-\Omega$ model of the solar dynamo}

\begin{abstract} We consider a conventional $\alpha-\Omega$-dynamo model 
with meridional circulation that exhibits typical
features of the solar dynamo, including a Hale cycle period of around 20 
years and a reasonable shape of the butterfly diagram. 
With regard to recent ideas of a tidal synchronization of the
solar cycle, we complement this model by an additional time-periodic
$\alpha$-term that is localized in the tachocline region. It is shown 
that amplitudes of some dm/s are sufficient for this $\alpha$-term 
to become capable of entraining the underlying dynamo. We argue that 
such amplitudes of $\alpha$ may indeed be realistic, since velocities 
in the range of m/s are reachable, e.g., for tidally excited 
magneto-Rossby waves.

\end{abstract}
\end{opening}
\section{Introduction}

The general idea that solar activity variations
might be linked to the orbital motion of the planets 
traces back to \cite{Wolf1859}, and 
was kept alive, throughout one and a half centuries, by a 
number of authors 
\citep{delarue1872,Bollinger1952,Jose1965,Takahashi1968,Wood1972,Opik1972,CondonSchmidt1975,Charvatova1997,Zaqa1997,Landscheidt1999,Palus2000,DeJager2005,Wolff2010,Abreu2012,Callebaut2012}.
The more specific coincidence, though, of the 11.07-year 
alignment cycle of the tidally dominant planets 
Venus, Earth and Jupiter with the Schwabe cycle was brought to the fore  
only recently by \cite{Hung2007,Scafetta2012,Wilson2013,Okhlopkov2016}.

Since even such a remarkable agreement between the average values of
two periods might still be a pure coincidence, the question of whether 
there is a  phase coherence between the two time series becomes of 
utmost importance. The possible phase stability of the Schwabe cycle 
was first discussed in the paper ``Is there a chronometer hidden 
deep in the Sun?''
by \cite{Dicke1978}. Analyzing the ratio between the mean
square of the {\it residuals} (i.e., the distances between 
the instants of the 
actual cycle maxima and the hypothetical maxima according to a linear 
trend) to the mean square of 
the differences between two consecutive residuals, Dicke's conclusions 
favoured a clocked process over a random walk process.
However, apart from the poor statistics connected with the mere 25 maxima 
taken into account, one should also take seriously Hoyng's 
later warning \citep{Hoyng1996}
that any $\alpha$-quenching mechanism could easily lead to  
a sort of self-stabilization of the 
solar dynamo, making a genuine random walk process ``disguise'' 
itself as a clocked process - at least for some centuries. 
A complementary type of cycle stability appears as a typical feature of 
conventional Babcock-Leighton
dynamos whose period is largely determined by the turnover time of the 
meridional circulation \citep{Dikpati1999,Charbonneau2000,Charbonneau2020}, 
which is indeed assumed to be much less fluctuating than 
the $\alpha$ effect in the convection zone.

With those caveats in mind, we had recently re-considered \citep{Stefani2020b} the 
longer time series of cycle minima/maxima as bequeathed to us by 
\cite{Schove1983}, and matched 
them with two series of the cosmogenic isotopes $^{10}$Be and $^{14}$C.
Apart from the hardly decidable existence, or not, of two ``lost cycles'' 
(or phase jumps) around 1563 \citep{Link1978}
and 1795 \citep{Usoskin2002}, our analysis confirmed, by and large, 
Dicke's conclusion in favour of a clocked cycle, 
now  throughout the last millennium. 
This result was then put into the context of the most remarkable, though
widely overlooked, work of \cite{Vos2004} whose analysis of two series 
of algae-related data from 10000-9000 cal. BP  had evidenced 
a phase-stable Schwabe cycle with a period of 11.04 years.

In view of those two independent thousand-year long segments 
showing nearly identical Schwabe cycles with average 
periods between 11.04 and 
11.07 years  and the strong evidence for phase stability in 
either case, we consider it at least worthwhile to quest for a 
possible physical mechanism that could be capable of linking 
the weak tidal forces as exerted by planets
with the solar dynamo. Setting out from the numerical 
observation 
\citep{Weber2013,Weber2015,Stefani2016}
that a tide-like influence
(with its typical $m=2$ azimuthal dependence) can entrain
the {\it helicity oscillation} of an underlying $m=1$ instability 
(the Tayler instability \citep{Tayler1973,Seilmayer2012} 
for that matter) with barely changing 
its energy content, we have 
pursued some rudimentary synchronization studies in the framework of 
simple 0D and 1D $\alpha-\Omega$-dynamo models
\citep{Stefani2017,Stefani2018,Stefani2019}.
Within the same framework, we recently tried \citep{Stefani2020a,Stefani2021} to 
explain also the longer term 
Suess-de Vries cycle in terms of a beat period \citep{Wilson2013,Solheim2013}
between the fundamental 22.14-year 
Hale cycle and the 19.86-yr period of the Sun's 
barycentric motion (forced, in turn, by the 
orbits of Jupiter and Saturn \citep{Cionco2018}) .
With the intervening spin-orbit coupling remaining poorly 
understood, we took resort to the  
same buoyancy instability mechanism as it had been been employed by 
\cite{Abreu2012}  to plausibilize 
typical modulation periods on the centennial time-scale. Yet, 
this similarity between the final results notwithstanding,
the fundamental time-scales of our model (22.14 and 19.86 years) 
that generate the much longer beat period of 193 years,  
are still close to the
period of the undisturbed dynamo. Our mechanism for explaining 
long-term modulations might, therefore, be
less vulnerable to stochastic noise than what was 
discussed by \cite{Charbonneau2022} in relation to the original model
of \cite{Abreu2012}.

Admittedly, being restricted to the latitudinal coordinate, our  
simple 1D dynamo model did not have the requisite level of detail 
to give a quantitative answer to  
Charbonneau's recent  question of 
``what, then, can be considered a physically 
reasonable amplitude for external forcing'' \citep{Charbonneau2022}. 
It was all the more encouraging that, utilizing a 2D Babcock-Leighton model 
with a periodic perturbation of the lower operating field threshold of the
source term,  \cite{Charbonneau2022} found a similarly robust
synchronization mechanism as \cite{Stefani2019}.
Such a variation of the lower operating field threshold   
would correspond to variations of the field loss term 
$\kappa$ as employed in \citep{Stefani2020a,Stefani2021} to parameterize
the spin-orbit coupling with its 19.86-yr periodicity.
While we do not exclude a viable physical translation of the
(11.07-yr periodic) tidal forcing into such a type of variation
of the field storage capacity,
in this paper we will stick to our original idea that
it is essentially the $\alpha$ effect that is affected by the 
tides. Specifically, we seek to know then {\it how much of this 
periodic $\alpha$ variation
would be needed} to accomplish synchronization of an otherwise  
conventional $\alpha-\Omega$-dynamo. Guided by a rough 
estimation based on the virial assumption
$U_{\rm pot}\approx E_{\rm kin}$, we consider approximately  1 m/s an upper limit 
for the tide-induced velocity variation. 
Given that the value of $\alpha$, which reflects only the helical 
part of the turbulence,
is typically one order of magnitude lower than
the underlying velocity, the focus of our modelling will be on whether 
$\alpha$-values of the order of dm/s are sufficient 
to entrain the entire solar dynamo.

To answer this very specific question, we step back from the more
sophisticated double-synchronization model of \cite{Stefani2020a,Stefani2021}
and restrict ourselves to the very basic tidal synchronization of
the Schwabe/Hale cycle. In the next section, we present a
rather conventional two-dimensional $\alpha-\Omega$-dynamo with
meridional circulation ${\bf u}_{\rm p}$, utilizing 
observation-constrained values
for $\Omega$ and ${\bf u}_{\rm p}$, and employing more or less 
realistic values of $\alpha$ and the magnetic 
diffusivity $\eta$.  To keep the model 
simple, no specific Babcock-Leighton source term is added to the 
$\alpha$-effect ``living'' in the convection zone. In the next section, 
we first adjust the value of 
$\eta$ to provide 
a reasonable natural period of the undisturbed dynamo. While the 
most simple form of the $\alpha-\Omega$ model leads, as usual, 
to a badly shaped butterfly diagram, the correct butterfly shape 
is recovered by
switching on the meridional circulation.
Based on the reference model thus defined, we will then assess 
in detail
how much $\alpha$ variation in the tachocline region is actually 
needed for synchronization.

The paper will conclude with a short discussion of the results and
some prospects for future work.

\section{The model}

In this section, we motivate and describe our mean-field solar dynamo model and 
discuss its numerical implementation. Considering only axi-symmetric solutions, 
we work with a system of partial 
differential equations whose spatial variables are
the co-latitude and the radius. Intentionally, the model has been 
kept similarly simple as the benchmark model of 
\cite{Jouve2008}.

As usual, the magnetic field is split into a poloidal component
${\bf{B}}_P(r,\Theta,t)=\nabla \times (A(r,\Theta,t) {\bf{e}}_{\phi})$
and a toroidal component ${\bf{B}}_T(r,\Theta,t)=B(r,\Theta,t) {\bf{e}}_{\phi}$.
The main sources of dynamo action are 
the gradient of the angular velocity $\Omega$ and
the $\alpha$-effect resulting from the helical part of the turbulence
in the convection zone.
While our model is not a Babcock-Leighton model (which would 
require a particular source term at the surface) 
it is a flux-transport model in that it comprises a meridional circulation 
${\bf u}_{\rm p}$, mainly to ensure a realistic shape of the butterfly 
diagram.

Choosing the solar radius $R_{\odot}=695700$\,km as the length and 
the diffusive time $R_{\odot}^2/\eta_t$ as the time scale,
we employ here - as in \cite{Jouve2008} - the dimensionless form of 
the coupled induction equations
for the azimuthal components $B:=B_{\phi}$ of the magnetic field
and $A:=A_{\phi}$ of the vector potential, 
\begin{eqnarray} 
  \frac{{\partial} B}{{\partial} t} &=& \tilde{\eta} D^2 B+ 
  \frac{1}{s}\frac{\partial (s B)}{\partial r}\frac{\partial \tilde{\eta}}{\partial r}
  - R_{\rm m} s {\bf u}_p \cdot \nabla \left( \frac{B}{s} \right)
  +C_{\Omega} s (\nabla \times (A{\bf{e}}_{\phi}))\cdot \nabla \Omega  \\
  \frac{{\partial} A }{{\partial} t} &=&\tilde{\eta} D^2 A- \frac{R_{\rm m}}{\rm s}{\bf u}_{\rm p} \cdot \nabla (s A)+
  C^c_{\alpha} \alpha^c B + C^p_{\alpha} \alpha^p B, 
    \label{systemequations}
   \end{eqnarray}
wherein we use the notations $D^2:=(\nabla^2-s^{-2})$, $s:=r \sin \theta$
and $\tilde{\eta}=\eta/\eta_{\rm t}$, with $\eta_{\rm t}$ being the turbulent 
magnetic diffusivity in the convection zone.

This system is governed by four magnetic Reynolds numbers characterizing,
respectively, the effects of shear, 
meridional circulation, and
two different $\alpha$ terms:
\begin{eqnarray} 
  C_{\Omega}&=&\Omega_{\rm eq} R^2_{\odot} /\eta_{\rm t} \label{reynolds1}\\
  R_{\rm m}&=&u_0 R_{\odot}/ \eta_{\rm t}    \label{reynolds2} \\
  C^c_{\alpha}&=&\alpha^c_{\rm max} R_{\odot} /\eta_{\rm t}  \label{reynolds3} \\
  C^p_{\alpha}&=&\alpha^p_{\rm max} R_{\odot} /\eta_{\rm t} \;.
    \label{reynolds4}
   \end{eqnarray}
Herein, $\Omega_{\rm eq}=2 \pi \times 456$\,nHz is the angular velocity at the equator,
and $u_0$ and $\alpha^c_{max}$  and $\alpha^p_{max}$ are the typical intensities of 
the meridional circulation and the 
two separate $\alpha$ effects in the convection zone and in the tachocline region. 
In contrast to \cite{Guerrero2007,Jouve2008,Sanchez2014}, 
we do not incorporate any specific Babcock-Leighton source term.

We suppose the turbulent magnetic diffusivity $\eta_{\rm t}$ in the convection zone
to be dominated by a strong $\beta$ effect, whereas it is much smaller
in the relatively quiet tachocline region. Refraining from more 
complicated structures of $\eta$ as employed, e.g., in \cite{Guerrero2007} or 
\cite{Sanchez2014}, we use here the simple form of \cite{Jouve2008}
   \begin{eqnarray} 
  \tilde{\eta}(r)&=& \frac{\eta_c}{\eta_t} +\frac{1}{2}\left(1- \frac{\eta_c}{\eta_t}\right) 
  \left[1 +{\rm erf} \left(\frac{r-r_c}{d}\right) \right]
  \label{eta}
   \end{eqnarray}
with $\eta_c=0.01 \eta_t$, $r_c=0.7$ and $d=0.02$, 
which shows a smoothed-out jump (by a factor of 100) between the 
radiation zone and the convection zone.

\begin{figure}[t]
  \centering
  \includegraphics[width=0.99\textwidth]{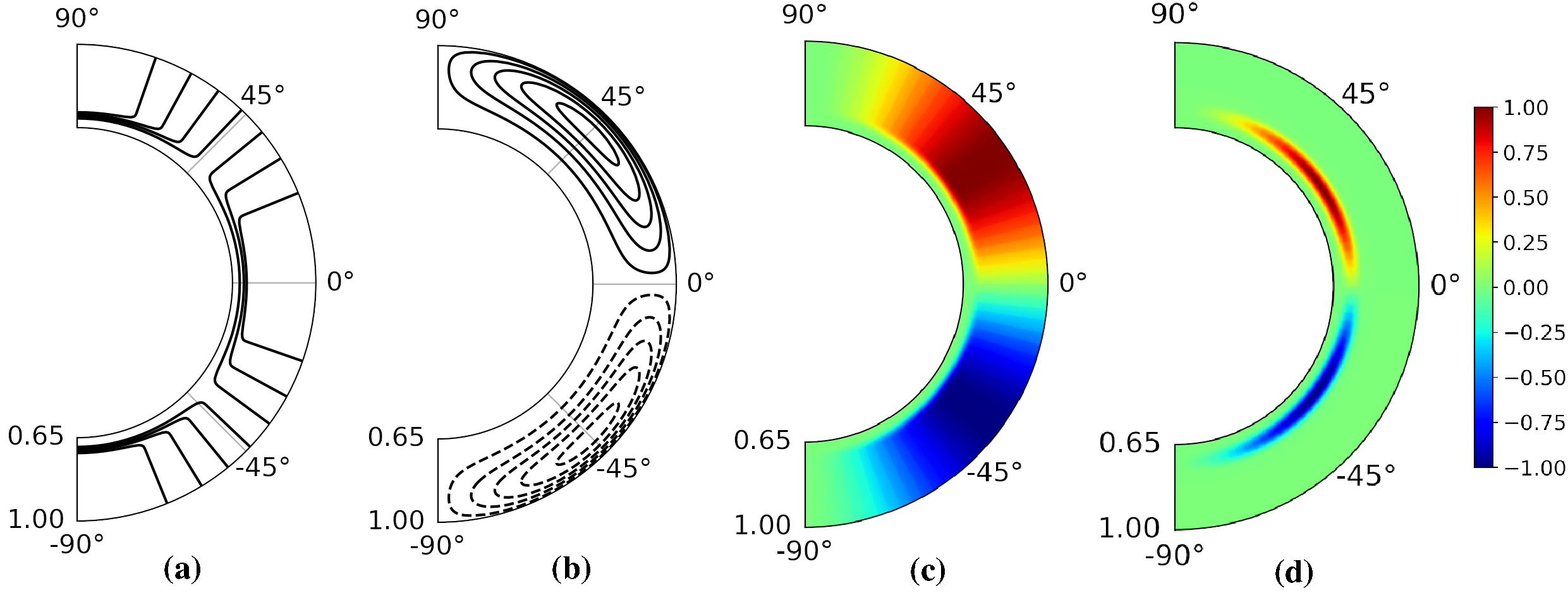}
  \caption{Spatial structures of the main ingredients of the dynamo model in the meridional plane. 
   (a) Isolines of $\Omega(r,\Theta)/\Omega_{\rm max}$.
  (b) Streamlines of ${\bf u}_{\rm p}(r,\Theta)$. 
  (c) Constant part of $\alpha$, taken in the unquenched state: $\alpha^c(r,\Theta)/\alpha^c_{\rm max}$.
  (d) Periodic part of $\alpha$, with the resonance term set to 1: $\alpha^p(r,\Theta)/\alpha^p_{\rm max}$.
  }
  \label{Fig:ingredients}
\end{figure}

For the angular velocity we apply the same spatial structure
as in \cite{Jouve2008}:
\begin{eqnarray} 
  {\Omega}(r,\Theta)&=&C_{\Omega} \left\{ \Omega_c+\frac{1}{2} \left[ 1 +{\rm erf} \left( \frac{r-r_c}{d} \right) \right]
  (1-\Omega_c-c_2 \cos^2 \Theta ) \right\}
  \label{omega}
   \end{eqnarray}
with $r_c=0.7$, $d=0.02$, $\Omega_c=0.92$ and $c_2=0.2$ (see Figure \ref{Fig:ingredients}(a)).
   
For the meridional circulation we chose, again as in \cite{Jouve2008},
one single cell defined by 
${\bf u}_{\rm p}= \nabla \times (\psi(r,\Theta) {\bf{e}}_{\phi})$ with
the stream function
 \begin{eqnarray} 
  \psi(r,\Theta)&=& R_{\rm m} \left\{ -\frac{2}{\pi} \frac{(r-r_b)^{2}}{(1-r_b)} \sin \left( \pi \frac{r-r_b}{1-r_b}  \right) 
  \cos \Theta \sin \Theta \right\}
  \label{meridional}
   \end{eqnarray}
with $r_b=0.65$ (see Figure \ref{Fig:ingredients}(b)). We are well aware
of the fact that the specific structure of ${\bf u}_{\rm p}$ is much less 
settled than that of ${\Omega}(r,\Theta)$, and that more complicated two-cell flows \citep{Kosovichev2022} might also be considered in future improvements of our model.

Finally,  $\alpha=\alpha^c+\alpha^p$ is thought to consist of a 
conventional part $\alpha^c$  in the convection zone, whose time-dependence 
stems only from the quenching  by the magnetic field,
\begin{eqnarray} 
  \alpha^c(r,\Theta,t)&=&C^c_{\alpha} \frac{3 \sqrt{3}}{4}\sin^2 \Theta \cos \Theta 
  \left[1+{\rm erf} \left( \frac{r-r_c}{d} \right)  \right]\left[1+\frac{|{\bf{B}}(r,\Theta,t)|^2}{B^2_0} \right]^{-1}
  \label{alphac}
   \end{eqnarray}
with $B_0=1$, and an explicitly  time-dependent (with forcing period $T_{f}$) part 
$\alpha^p$ that is concentrated in the tachocline region,
    \begin{eqnarray} 
  \alpha^p(r,\Theta,t)&=&C^p_{\alpha} \frac{1}{\sqrt{2}}\sin^2 \Theta \cos \Theta 
  \left[1+{\rm erf} \left( \frac{r-r_c}{d} \right)  \right]
  \left[1-{\rm erf} \left( \frac{r-r_d}{d} \right)  \right] \times \nonumber \\  
  &&  \times \frac{|{\bf{B}}(r,\Theta,t)|^2}{1+|{\bf{B}}(r,\Theta,t)|^4}  \sin(2 \pi t/T_f  )\;,
  \label{alphap}
   \end{eqnarray}
where $r_d=0.75$. Note that the factor on the second line of Eq. (\ref{alphap}) represents a 
resonance term as introduced in \cite{Stefani2016} in order to account for a
field-dependent optimal reaction of the underlying instability (e.g., 
Tayler instability) on the tidal forcing. 
A similar field dependence has been used, e.g., in 
\cite{Charbonneau2022}, although with the slightly different 
interpretation as a nonlinearity of the non-local source 
term that incorporates both a lower and upper operating threshold 
on the strength of the toroidal magnetic at the base of the convection zone.
The spatial structures of these two $\alpha$ terms are visualized in 
Figure \ref{Fig:ingredients}(c,d), in either case disregarding any magnetic-field 
dependence.
   
For the numerical solution, an explicit finite difference scheme in 
two dimensions in spherical coordinates is used. As in 
\cite{Ruediger2003}, 
the standard resolution
was $64 \times 64$ grid points in both radial and latitudinal directions.
The equations are solved with perfect conductor boundary conditions
$A=\partial (r B)/ \partial r=0$ at $r=0.65 R_{\odot}$ and vertical 
field conditions $B_{\phi}=B_\Theta=0$ at $r=R_{\odot}$.

\section{Results}

In this section we present and assess the results of  three dynamo models with 
increasing complexity. 

\subsection{Non-synchronized model, without meridional circulation}

\begin{figure}[t]
  \centering
  \includegraphics[width=0.99\textwidth]{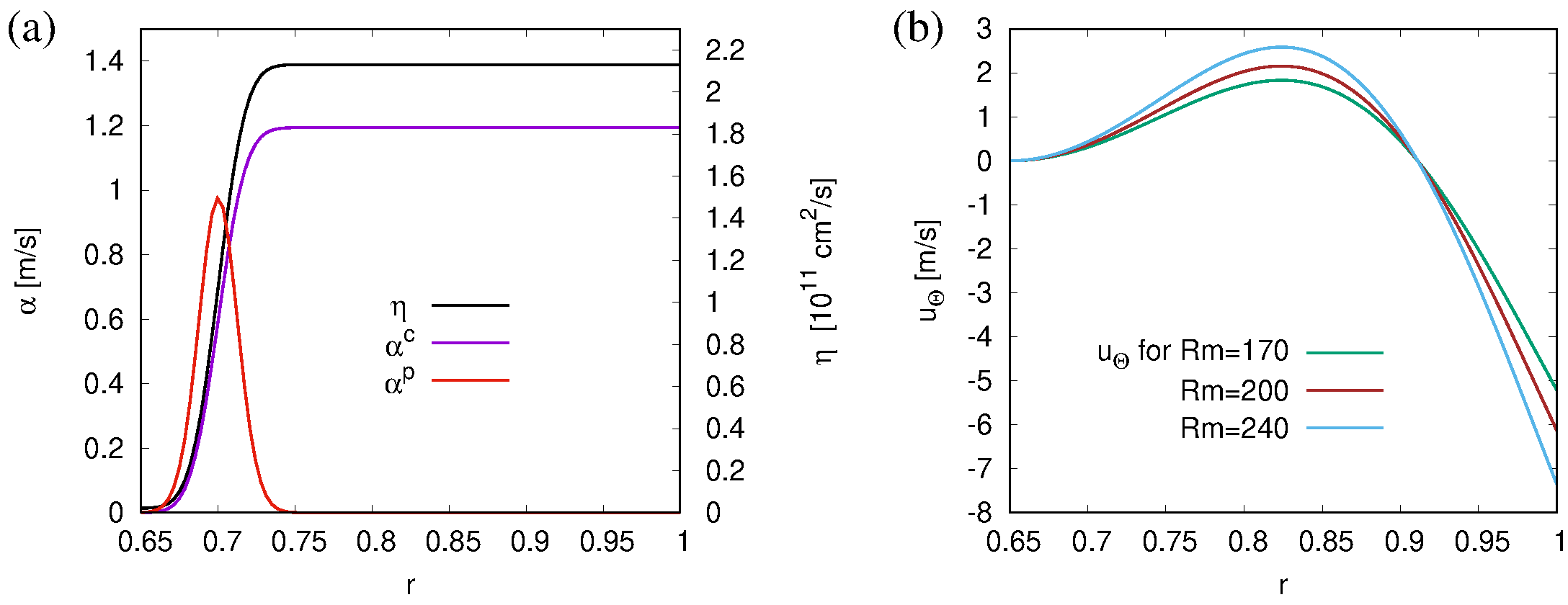}
  \caption{Radial dependence of various dynamo ingredients in physical units, all taken 
  at $\Theta=45^{\circ}$.
  (a) Diffusivity $\eta(r)$ (black), $\alpha^c(r)$ in the unquenched form (violet),
  and $\alpha^p(r)$ for $\alpha^p_{\rm max}=\alpha^c_{\rm max}$ and 
  with the field-dependent resonance factor artificially set to 1 (red). 
  (b) $u_\Theta(r)$ resulting from the stream function of Eq. (9) for three 
  different $R_m$.
  }
  \label{Fig:alphaetameri}
\end{figure}

First we consider the simplest case of a Parker's migratory dynamo \citep{Parker1955}, 
without any synchronization term ($\alpha^p=0$), 
and without meridional circulation (${\bf u}_{\rm p}=0$).
For the sake of concreteness, we set $\eta_t=2.13 \times 10^{11}$\,cm$^2$/s, and
$\alpha^c_{\rm max}=1.30$\,m/s, which both are close to the respective 
geometric means of the lower and upper values as typically 
found in the literature ($10^{10}-10^{13}$\,cm$^2$/s for $\eta$ and 
$10-10^{3}$\,cm/s for $\alpha$, see  \cite{Charbonneau2020}).  The resulting 
magnetic Reynolds numbers according to 
Equations (\ref{reynolds1}) and (\ref{reynolds3}) are 
$C_{\Omega}=65100$ and $C^c_{\alpha}=42.46$. 
The radial dependencies of $\eta(r)$ and
$\alpha^c$ (in its unquenched form) are illustrated, for $\Theta=45^{\circ}$,
in Figure  \ref{Fig:alphaetameri}(a). Note that at this particular 
angle $\alpha^c(r)$ does not 
reach the maximum value of $1.30$\,m/s.

\begin{figure}[t]
  \centering
  \includegraphics[width=0.99\textwidth]{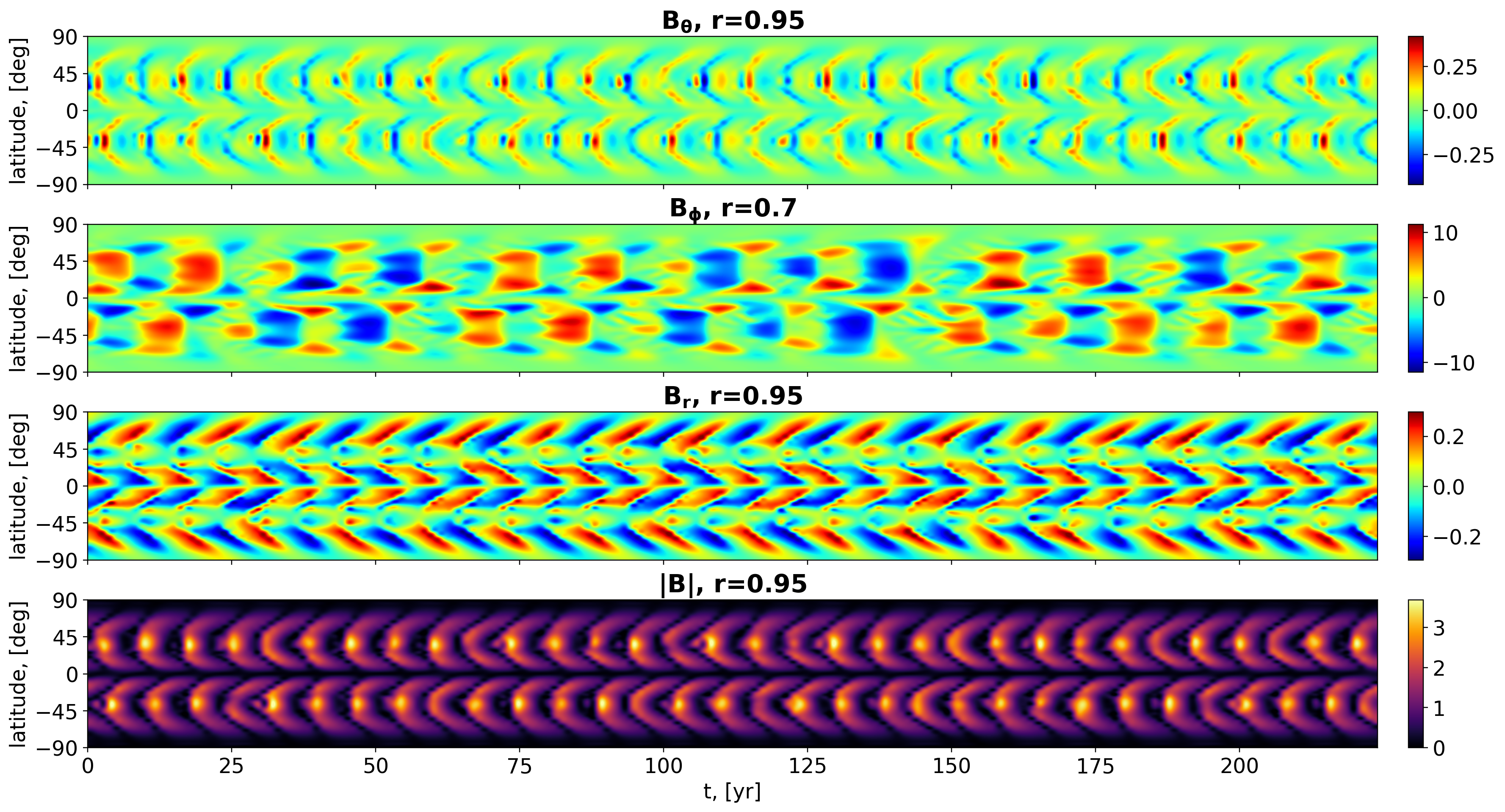}
  \caption{Contour-plots ${\bf B}_{\Theta}(r=0.95,\theta,t)$, 
  ${\bf B}_{\phi}(r=0.7,\theta,t)$, and ${\bf B}_{r}(r=0.95,\theta,t)$
  and of 
   $|{\bf B}(r=0.95,\theta,t)|$ for the 
  non-synchronized model without
  meridional circulation. 
 Note that the ordinate axis represents not the  
 colatitude $\theta$ but the normal solar latitude  $90^{\circ}-\theta$.
  }
  \label{Fig:nosynchronomeri}
\end{figure}

Figure \ref{Fig:nosynchronomeri} illustrates the resulting field dependence on 
time and latitude, taken at $r=0.95$, showing a reasonable 
dynamo cycle period of $T_d=14.27$\,years (i.e. 0.0198 diffusion times), but a 
badly shaped butterfly 
diagram.

\subsection{Non-synchronized model, with meridional circulation}

In order to recover the correct shape of the
butterfly diagram, we switch on a meridional circulation, setting its value to
$u_0=5.2$\,m/s, which corresponds to $R_{\rm m}=170$. 
For this value, as well as for $R_{\rm m}=200$ and 240, 
the radial dependence of $u_{\Theta}$
is shown, again for $\Theta=45^{\circ}$,
in Figure  \ref{Fig:alphaetameri}(b). 
While the values $u_{\Theta}$ at $r=1$ are by factor of appr. two too 
low compared with observations, the typical values of 1-2 m/s at 
the base of the convection zone 
are quite compatible with values from helioseismology. 
Actually, the latter velocities are the crucial ones to set the cycle
period.

\begin{figure}[t]
  \centering
  \includegraphics[width=0.99\textwidth]{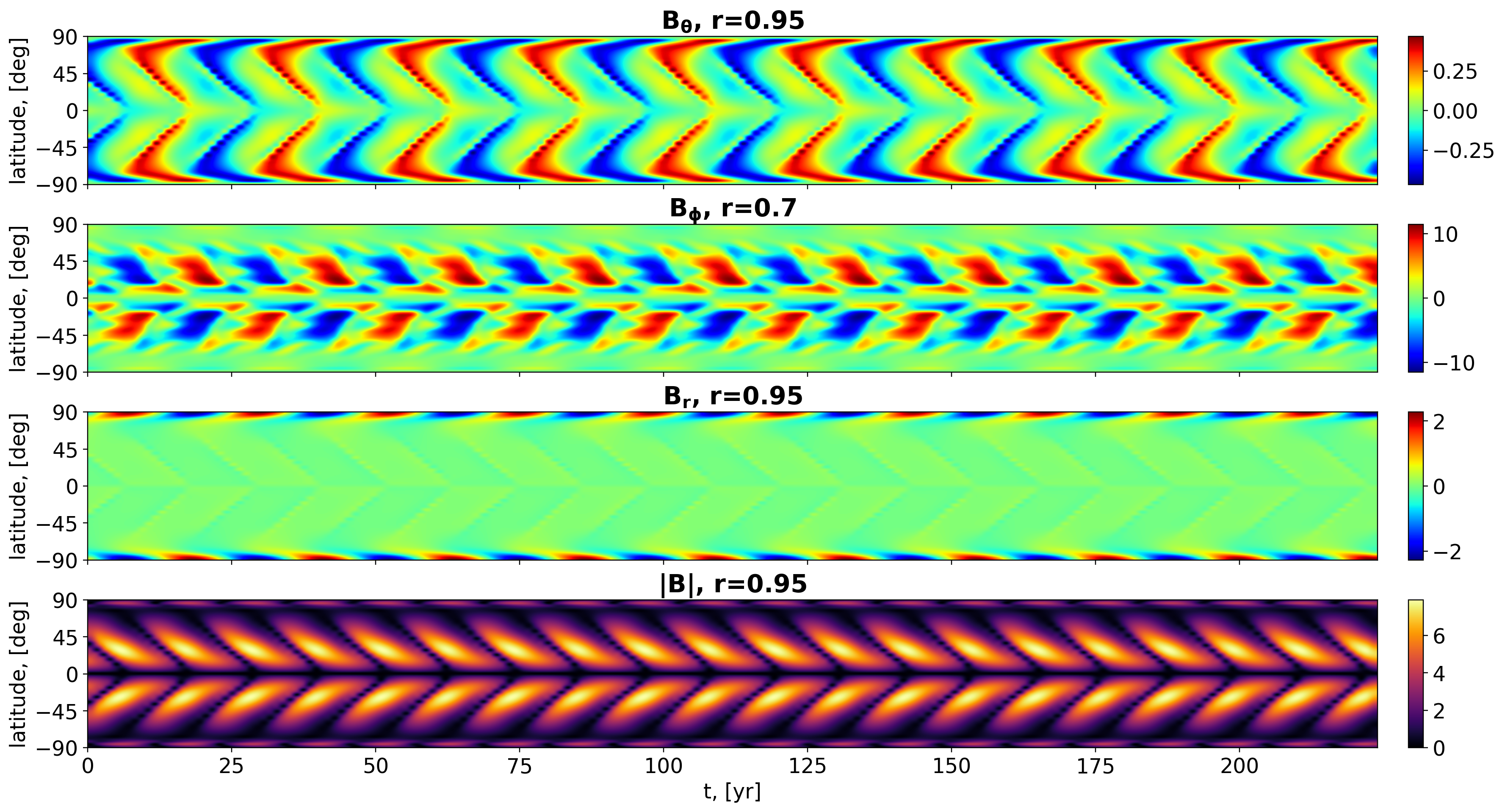}
  \caption{Same as Figure \ref{Fig:nosynchronomeri}, but including meridional 
  circulation with $R_{\rm m}=170$.
  }
  \label{Fig:nosynchromeri}
\end{figure}

As seen in Figure \ref{Fig:nosynchromeri}
we obtain now
a butterfly diagram of rather decent shape, and a slightly changed cycle period 
of $T_d=22.798$\,years. This  will serve in the following as the reference dynamo 
model whose synchronization is to be evaluated thereupon. 
While further improvements of the spatio-temporal features
of the magnetic field 
are certainly possible (for example, when including an
appropriate Babcock-Leighton source term), we refrain from any further sophistication of the model.

\subsection{Synchronized model}

Finally, we switch on the periodic $\alpha$ term with an assumed forcing 
period of  $T_f=11.00$\,years (we do not insist here on the precise value
of 11.07 years). The radial dependence of $\alpha^p$ 
is illustrated by the red curve in Figure  \ref{Fig:alphaetameri}(a). Note, 
however, that here $\alpha^p_{\rm max}$ has the same value of $1.30$\,m/s
as the corresponding $\alpha^c_{\rm max}$, and that the field-dependent resonance term in 
Equation (11) is artificially set to 1.
In reality, the resonance term reduces this 
value by a factor of 2 for the optimum field 
strength, and even more so outside the optimum.

\begin{figure}[t]
  \centering
  \includegraphics[width=0.99\textwidth]{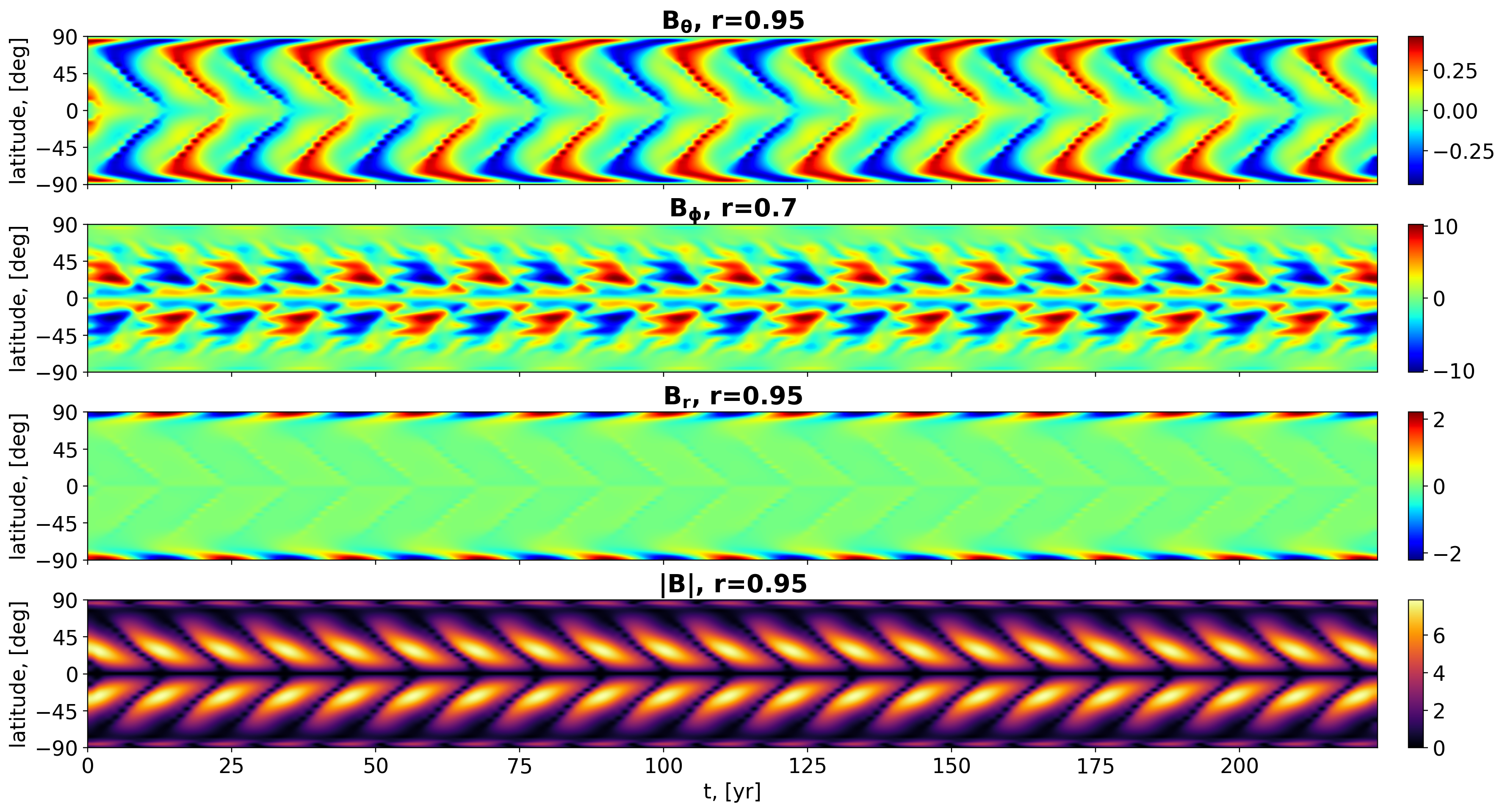}
  
  \caption{Same as Figure \ref{Fig:nosynchromeri}, but with synchronization by a 
  periodic $\alpha$-term
  with amplitude $\alpha^p_{\rm max}=0.52$\,m/s and period  $T_f=11.00$\,years.
  }
  \label{Fig:synchromeri}
\end{figure}

As shown in Figure \ref{Fig:synchromeri}, for the specific value
$\alpha^p_{\rm max}=0.52$\,m/s we obtain now the dynamo period $T_d=22.00$\,years
which corresponds to twice the period $T_f$ of 
the forcing.

\begin{figure}[t]
  \centering
  \includegraphics[width=0.99\textwidth]{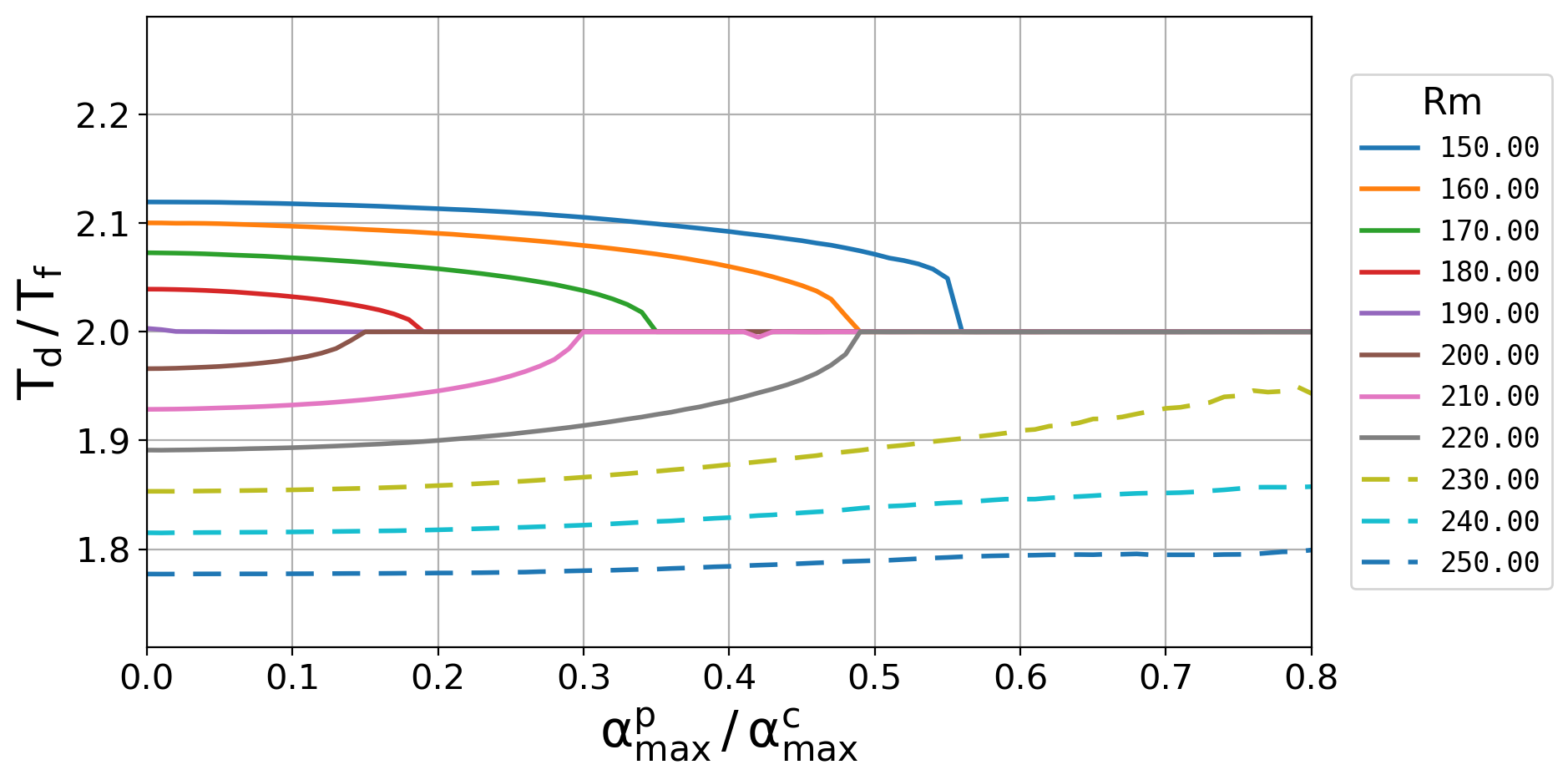}
  \caption{Ratio of the period $T_d$ of the signal  
  to the period $T_f$ of the forcing 
  in dependence on the relative strength of the forcing 
  $\alpha^p_{\rm max}/\alpha^c_{\max}$. The color coded curves
  refer to different ratios of the ``natural'' 
  period $T_n$ of the non-synchronized dynamo to $T_f$,  
  which has been varied by
  changing the magnetic Reynolds number $R_m$ 
  of the meridional circulation. $T_n$ can be read off from 
  the value at the ordinate axis multiplied by 11\,years; it amounts, 
  for example, to 23.3\,years for $R_m=150$, to 21.6\,years 
  for $R_m=200$, 
  and to 19.5\,years for $R_m=250$.
  }
  \label{Fig:synchro1}
\end{figure}

In Figure \ref{Fig:synchro1} we plot the dependence of the dynamo period $T_d$ 
on $\alpha^p_{\rm max}$. Here we have used a couple of ratios of the
``natural'' period $T_n$ (of the non-synchronized dynamo with 
  $\alpha^p_{\rm max }=0$) to the forcing periods 
$T_f$ by simply changing the amplitude of meridional circulation which governs  $T_n$.
Very similar to Figure 10 in \cite{Stefani2019}, and to Figure 10 in
\cite{Charbonneau2022}, we obtain a clear parametric resonance for some
critical value of $\alpha^p_{\rm max}$ that depends on the initial distance between
twice the forcing period $T_f$ and the natural period $T_n$ of the unperturbed dynamo.
As we had chosen $\alpha^c_{\rm max}=1.30$\,m/s, synchronization occurs for
an amplitude of $\alpha^p_{\rm max}$ in the range of some dm/s.
The relative smallness of this number is, of course, a consequence of the 100 times smaller
value of $\eta$ in the tachocline region which amplifies correspondingly the 
induction effect of $\alpha^p$, even if the latter 
is concentrated in a significantly 
smaller zone than $\alpha^c$.
That said, we must also admit that synchronization
requires a certain proximity of $2 T_f$ and $T_n$; for 
the $R_m$ values indicated by the dashed lines in Figure \ref{Fig:synchro1}
no clear synchronization was 
observed even for the highest considered value of $\alpha^p_{\rm max}/\alpha^c_{\rm max}=1$.
This narrowness of the synchronizability region, which somewhat contrasts with 
the broader region obtained in frame of the 1D model (Fig. 10 of \cite{Stefani2019}), 
might have to do with the 
tight scaling of $T_n$ with the period of the
meridional circulation.

\section{Conclusions}

As a sequel to the 0D and 1D modelling of solar cycle 
synchronization \citep{Stefani2016,Stefani2018,Stefani2019,Stefani2020a,Stefani2021}, 
we have investigated a more realistic 2D $\alpha-\Omega$-dynamo model.
Starting from a conventional set-up without meridional circulation,
exhibiting a badly shaped butterfly diagram, via 
an enhanced model with meridional circulation showing the correct
butterfly shape, we have assessed the
synchronization capabilities of a time-periodic 
$\alpha$ term concentrated in the tachocline region.
For rather standard values of all other parameters, 
it was shown that synchronization starts already for a magnitude
of this additional $\alpha$-term as low as some dm/s. The smallness 
of this value relies on the fact 
that $\eta$ in the quiet tachocline region is significantly lower 
than in the convection zone where it is dominated by the 
turbulent $\beta$ effect. The utilized tachoclinic 
diffusivity $\eta \approx 2.13 \times 10^9$\,cm$^2$/s should be 
considered a conservative choice;
in view of much lower values such as, e.g. $2.2 \times 10^8$\,cm$^2$/s
as used by \cite{Guerrero2007}, the real value of $\alpha$, 
required for synchronization, might still be lower than the one derived here.

This brings us back to Charbonneau's ``elephant in the room: what, then,
can be considered a physically reasonable amplitude for external forcing?''
\citep{Charbonneau2022}. 
Let us recall the very rough energetic consideration 
\cite{Opik1972} that the typical tidal height of
$h_{\rm tidal} = G m R^2_{\rm tacho}/(g_{\rm tacho} d^3) \approx 1$\,mm  
corresponds energetically to a velocity scale 
of $v_0 \sim (2 g_{\rm tacho} h_{\rm tidal})^{1/2} \approx 1$\,m/s
when employing the huge gravity at the tachocline of 
$g_{\rm tacho} \approx 500$\,m/s$^2$. 
Invoking the  equally rough estimate $\alpha \sim v_0$ 
from renormalization theory \citep{Moffatt2019} (and even when
realistically assuming $\alpha$ to be one or two orders of 
magnitude smaller than $v_0$), a tidally generated 
$\alpha$-value 
of a few dm/s seems not out of reach.
Indeed, it was recently shown \citep{Gerrit} that 
(magneto-)Rossby waves \citep{Marquez2017,Zaqarashvili2018,Dikpati2020} 
under the influence of a {\it realistic} 
tidal forcing are capable of acquiring 
velocity scales of up to 1\,m/s.
Therefore, it appears that the ``astrological homeopathy''
\citep{Charbonneau2022} of tidal forcing may well be suited to 
generate an $\alpha$-effect in the tachocline region 
that is strong enough to entrain the entire solar dynamo. 

We have further confirmed 
the prior results of \cite{Stefani2019} (Figure 10) and \cite{Charbonneau2022} 
(Figure 10) that this type of synchronization requires 
a certain proximity of the tidal forcing's period 
to half the ``natural'' period of the undisturbed dynamo.  
The Sun, therefore, may just be in the 
lucky situation of being orbited by a Jupiter with a
period that fits nicely to half the ``natural'' 
period of the undisturbed dynamo.
It remains to be seen whether some peculiar features of the solar dynamo,
e.g its somewhat unusual cycle period 
\citep{Boehm2007} and, in particular,  
``its comparatively smooth, regular activity cycle''
\citep{Radick2018}, could find an explanation at this point.

What are the next steps to be taken? First and foremost, 
the specific action of $m=2$ tidal forces on various $m=1$
instabilities (e.g., Tayler) or waves (e.g., magneto-Rossby), 
and on the $\alpha$ effect connected with them, 
has to be quantified in a reliable manner. Complementary work on
tidal influences on Rayleigh-B\'enard convection, and
its large-scale circulation
\citep{Stepanov2019,Juestel2020,Juestel2022},
might be helpful to elucidate 
helicity entrainment in a more generic sense.

Second, the possible role of further axisymmetric induction 
effects, 
beyond the $\alpha$ effect,
has to be clarified. The basic idea of 
a torque-influenced magnetic buoyancy instability
within the tachocline
\citep{Ferrizmas1994,Zhang2003,Abreu2012} might play a central 
role here. It was indeed
employed as the basic synchronization
mechanism by \cite{Charbonneau2022}, while in \cite{Stefani2020a,Stefani2021},
we had used it only to bring into play the second fundamental 
period 19.86 years via spin-orbit coupling (yet 
poorly understood, but see \cite{Javaraiah2003,Shirley2006,Sharp2013} for first estimates).
It certainly needs
much more work to disentangle these two effects.
Further to this, we should not overlook alternative 
axisymmetric ($m=0$) 
instabilities, the possible relevance of which 
had been discussed by several authors 
\citep{Dikpati2009,Rogers2011}. The recently discovered
helical magnetorotational instability 
for flows with positive radial shear  
\citep{Mama2019} might be an particularly interesting 
candidate in this respect.

\begin{acks}
This work received funding from the European Research Council (ERC) under the 
European Union's Horizon 2020 research and innovation programme
(grant agreement No 787544).
We are grateful to Detlef Elstner for providing us with 
the finite-difference dynamo code. 
F.S. would like to thank 
J\"urg Beer, Robert Cameron, Antonio Ferriz Mas, Peter Frick, 
Gerrit Horstmann, Henri-Claude Nataf, Rafael Rebolo,
G\"unther R\"udiger, Dmitry Sokoloff, Willie Soon, Steve Tobias, 
Rodion Stepanov, 
Tom Weier, Ian Wilson and Teimuraz Zaqarashvili 
for helpful discussions on various aspects of the 
solar dynamo and its possible synchronization.
 L.J. acknowledges support from the Institut Universitaire de France.
\end{acks}

\section*{Disclosure of Potential Conflicts of Interest}
The authors declare that they have no conflicts of interest.

\end{article} 

\end{document}